\documentclass[preprint]{aastex62}

\usepackage{amsmath}
\usepackage{rotating}
\usepackage{comment}
\usepackage{listings}

\graphicspath{{./}{figures/}}

\accepted{April 12, 2020}

\shorttitle{NGC~3256 recombination lines}
\shortauthors{Michiyama et al.}

\begin{document}

\title{Star formation traced by optical and millimeter hydrogen recombination lines and free-free emissions in the dusty merging galaxy NGC~3256\\
-- MUSE/VLT and ALMA synergy --}

\correspondingauthor{Tomonari Michiyama}
\email{t.michiyama.astr@gmail.com}

\author{Tomonari Michiyama}
\affiliation{Kavli Institute for Astronomy and Astrophysics, Peking University, 5 Yiheyuan Road, Haidian District, Beijing 100871, P. R. China}
\affiliation{Department of Astronomical Science, SOKENDAI (The Graduate University for Advanced Studies), 2-21-1 Osawa, Mitaka, Tokyo 181-8588}
\affiliation{National Astronomical Observatory of Japan, National Institutes of Natural Sciences, 2-21-1 Osawa, Mitaka, Tokyo, 181-8588}

\author{Daisuke Iono}
\affiliation{National Astronomical Observatory of Japan, National Institutes of Natural Sciences, 2-21-1 Osawa, Mitaka, Tokyo, 181-8588}
\affiliation{Department of Astronomical Science, SOKENDAI (The Graduate University for Advanced Studies), 2-21-1 Osawa, Mitaka, Tokyo 181-8588}

\author{Kouichiro Nakanishi}
\affiliation{National Astronomical Observatory of Japan, National Institutes of Natural Sciences, 2-21-1 Osawa, Mitaka, Tokyo, 181-8588}

\affiliation{Department of Astronomical Science, SOKENDAI (The Graduate University for Advanced Studies), 2-21-1 Osawa, Mitaka, Tokyo 181-8588}

\author{Junko Ueda}
\affiliation{National Astronomical Observatory of Japan, National Institutes of Natural Sciences, 2-21-1 Osawa, Mitaka, Tokyo, 181-8588}
\affiliation{Harvard--Smithsonian Center for Astrophysics, 60 Garden Street, Cambridge, MA 02138, USA}

\author{Toshiki Saito}
\affiliation{Max-Planck-Institut f\"ur Astronomie, K\"onigstuhl 17, 69117 Heidelberg, Germany}

\author{Takuji Yamashita}
\affiliation{National Astronomical Observatory of Japan, National Institutes of Natural Sciences, 2-21-1 Osawa, Mitaka, Tokyo, 181-8588}
\affiliation{Research Center for Space and Cosmic Evolution, Ehime University, 2-5 Bunkyo-cho, Matsuyama, Ehime 790-8577, Japan}

\author{Alberto Bolatto}
\affiliation{Department of Astronomy and Laboratory for Millimeter-Wave Astronomy, University of Maryland, College Park, MD 20742, USA}

\author{Min Yun}
\affiliation{Department of Astronomy, University of Massachusetts, Amherst, MA 01003, USA}



\begin{abstract}
A galaxy--galaxy merger and the subsequent triggering of starburst activity are fundamental processes linked to the morphological transformation of galaxies and the evolution of star formation across the history of the Universe.
Both nuclear and disk-wide starbursts are assumed to occur during the merger process. 
However, quantifying both nuclear and disk-wide star formation activity is non-trivial because the nuclear starburst is dusty in the most active merging starburst galaxies.
This paper presents a new approach to this problem: combining hydrogen recombination lines in optical, millimeter, and free--free emission. Using NGC~3256 as a case study, H$\beta$, H40$\alpha$, and free--free emissions are investigated using the Multi Unit Spectroscopic Explorer at the Very Large Telescope of the European Southern Observatory (MUSE/VLT) and the Atacama Large Millimeter/submillimeter Array (ALMA). 
The H$\beta$ image obtained by MUSE identifies star-forming regions outside the nuclear regions, suggesting a disk-wide starburst. In contrast, the H40$\alpha$ image obtained by ALMA identifies a nuclear starburst where optical lines are undetected due to dust extinction ($A_{\rm V}\sim25$). 
Combining both MUSE and ALMA observations, we conclude that the total SFR is $49\pm2~M_{\odot}$~yr$^{-1}$ and the contributions from nuclear and disk-wide starbursts are $\sim34~\%$ and $\sim66~\%$, respectively. This suggests the dominance of disk-wide star formation in NGC~3256.
In addition, pixel-by-pixel analyses for disk-wide star-forming regions suggest that shock gas tracers (e.g., CH$_3$OH) are enhanced where gas depletion time ($\tau_{\rm gas}$=$M_{\rm gas}/SFR$) is long. This possibly means that merger-induced shocks regulate disk-wide star formation activities.
\end{abstract}

\keywords{galaxies: individual (NGC~3256) --- galaxies: interactions --- galaxies: irregular --- galaxies: starburst  --- submillimeter}


\section{Introduction} \label{sec:intro}
It has been known for decades that mergers of two disk galaxies can induce “nuclear starbursts” in the central $\sim1$~kpc region during the coalescence stage \citep[e.g.,][]{Keel_1985}, triggered by massive gas inflows.
In a more early stage of a merger, ``disk-wide starbursts"  ($\gtrsim1$~kpc) are seen both in theoretical models \citep{Barnes_2004} 
e.g., the Antennae galaxy \citep{Wang_2004}, Arp~140 \citep{Cullen_2007}, and NGC~2207+IC~2163 \citep{Elmegreen_2005}.
In particular, \citet{Cortijo-Ferrero_2017} investigated the star formation history of merging galaxies using optical integral field unit (IFU) observations, showing that disk-wide starbursts arise in the early stages whereas nuclear starbursts occur in the more advanced stages of a merger process.
Theoretical models predict that such disk-wide starbursts can be explained by interstellar medium (ISM) turbulence and fragmentation into dense clouds in the disk region \citep{Teyssier_2010, Bournaud_2011}. 
In addition, \citet{Saitoh_2009} suggest that shock-induced star formation may be efficient during a merger process.
Observationally, it is difficult to quantify both nuclear and disk-wide starbursts in a consistent manner.
For example, the mapping of hydrogen recombination lines (i.e., H$\alpha$ and  H$\beta$) by optical IFUs enables us to investigate the spatial distributions of star formation activities in regions where dust extinction is insignificant  \citep[e.g.,][]{Thorp_2019, Pan_2019} such as the disk component of galaxies.
However, optical observation is hampered by extinction from thick layers of interstellar dust clouds, and correct quantification of the star formation activity in dusty regions such as the central nucleus of a merging galaxy is highly non-trivial. 
One of the best methods of investigating the properties of star formation activities in such extremely dusty regions is hydrogen recombination lines in the millimeter (mm) range \citep{Scoville_2013}.
Recently, Atacama Large Millimeter/- submillimeter Array (ALMA) has detected recombination lines from nearby galaxies; e.g., NGC~253 \citep{Bendo_2015}, NGC~4945 \citep{Bendo_2016}, and NGC~5253 \citep{Bendo_2017}.
By cross-checking star formation rate (SFR) measurements from the other wavelengths, \citet{Bendo_2015, Bendo_2016, Bendo_2017} demonstrated that ALMA is effective to study the starburst activity in dusty regions ($A_{\rm V}\gtrsim$10). In this paper, we apply this method to investigate dusty starbursts in a merging galaxy. 

The SFR estimated from the hydrogen recombination line luminosity (hereafter, SFR$_{\rm RL}$) allows us to estimate the calibration constant between SFR and total infrared (TIR) luminosity. 
The calibration constant changes depending on the duration of the currently observed starbursts  \citep{Calzetti_2013}, because both high-mass short-lived stars and low-mass long-lived stars heat the dust and contribute to the TIR emission.
If a young stellar population is the predominant energy source within a system, the TIR emission is mainly produced by dust heated by these young stars. 
However, the recombination line mainly traces the current starbursts because only stars more massive than $\sim20~M_{\sun}$ produce a measurable ionizing photon flux.
As such, the ratio between SFR$_{\rm RL}$ and recombination line luminosity is constant (when the age of starburst is longer than $\sim6$~Myr.), allowing us to estimate the age of the starburst by comparing SFR$_{\rm RL}$ with the TIR luminosity. 
Hence, if the age of the starburst is shorter than the age of the galaxy merger, it is likely that the starburst was triggered by the galaxy interaction. 

Little has been reported on the observations of mm recombination lines in merging galaxies. 
For example, in the case of Arp 220, \citet{Anantharamaiah_2000} detected H42$\alpha$, H40$\alpha$, and H31$\alpha$ using IRAM 30m telescopes, and the results suggest multiple starbursts.
\citet{Scoville_2015} searched for H26$\alpha$ emission from Arp 220, but detection was unclear due to the contamination of a nearby HCN(4--3) line.
In order to investigate optical and millimeter hydrogen recombination lines, we focus on one specific merging galaxy, NGC 3256.
In this galaxy, H40$\alpha$ and H42$\alpha$ were detected by ALMA \citep{Harada_2018} and \citet{Erroz-Ferrer_2019, denBrok_2020} mapped the H$\alpha$ and H$\beta$ emissions with the Multi Unit Spectroscopic Explorer at the Very Large Telescope of the European Southern Observatory (MUSE/VLT)  \citep{Bacon_2010}.

NGC~3256 (redshift z = 0.00935\footnote{The redshift is from the NASA/IPAC Extragalactic Database (NED) (\url{https://ned.ipac.caltech.edu}).}) is a merging galaxy with a TIR luminosity (5--1100$\mu$m, $L_{\rm TIR}$) of  $4.8\times10^{11}~{L}_{\sun}$ (see SECTION~\ref{TIR} in detail)\footnote{We have adopted $H_0 = 67.7$~km~s$^{-1}$~Mpc$^{-1}$ and $\Omega_m = 0.307$ \citep{Planck_2016} as cosmological parameters throughout this article.}.
This system is at a distance of $D \sim$ 41.7~Mpc, which translates to 1$\arcsec\sim$198 pc.
There are two nuclei (northern and southern) separated by $\sim$  970 pc in NGC~3256.
The systematic velocity of the merger is assumed to be $cz\sim2800$~km~s$^{-1}$ ($c$ is light speed).
\citet{Lira_2002, Lira_2008} derive normalization of the extinction curve ($A_{\rm V}=5.5$ and $16$ for the northern and southern nuclei, respectively) from the {\it NICMOS}~\footnote{The Near Infrared Camera and Multi-Object Spectrometer (NICMOS) mounted on the Hubble Space Telescope (HST)} $H-K$ color.
The large $A_{\rm V}$ makes it impossible to investigate the southern nuclear starburst activity using optical hydrogen recombination lines. 
The southern nucleus is an ideal laboratory to quantify how much the H$\alpha$ and H$\beta$ emission miss the SFR using H40$\alpha$ emission.
In SECTION  \ref{Data}, the VLT and ALMA observations are explained. In SECTION  \ref{Analysis}, the formula to calculate the SFR is introduced. In SECTION  \ref{Discussion}, we investigate the nuclear starbursts, disk-wide starbursts, starburst timescale, and electron temperature. Finally, we summarize this project in SECTION  \ref{Summary-sec}.

\color{black}
\section{Data}\label{Data}
\subsection{MUSE}\label{sec:MUSE}
\begin{figure*}[!htbp]
\begin{center}
\includegraphics[width=18cm]{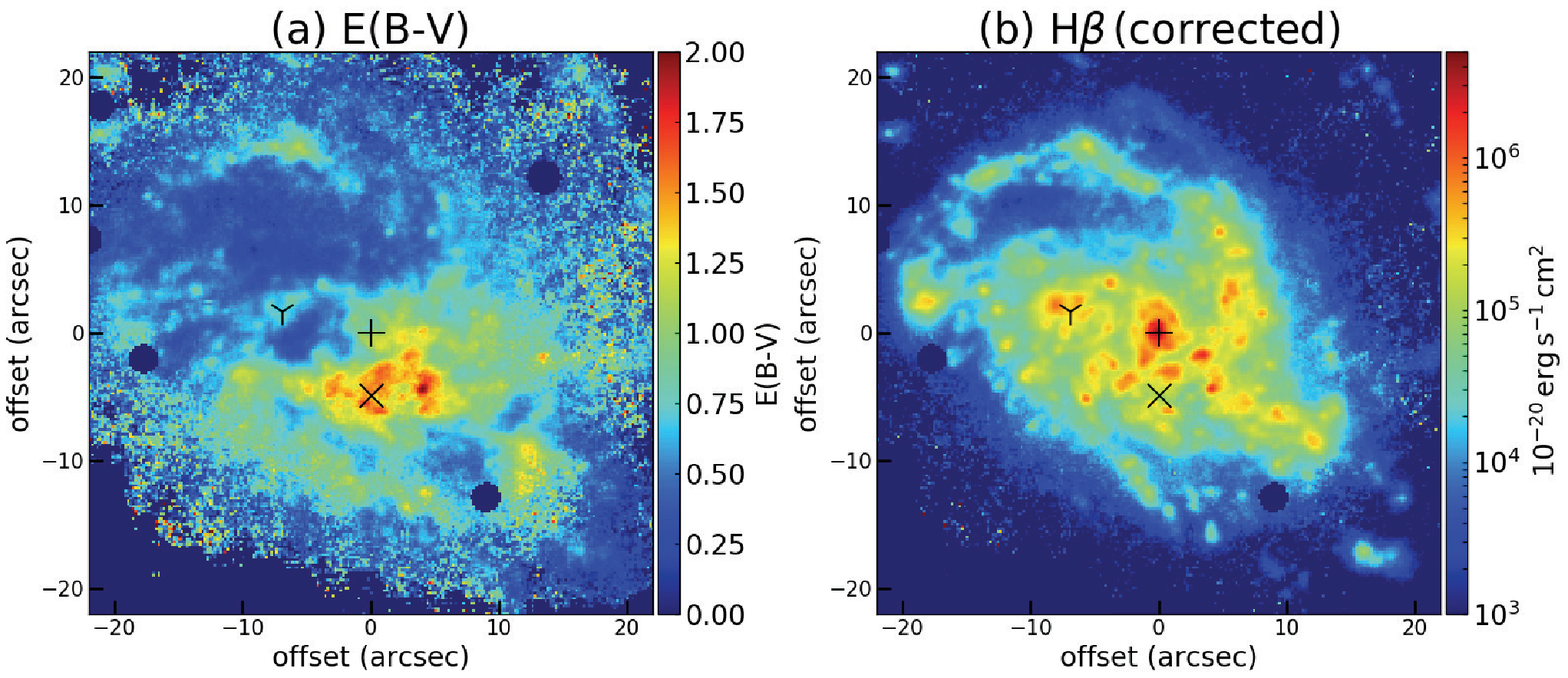}
\caption{(a) Extinction map and (b) dust-extinction-corrected  H$\beta$ map obtained by MUSE. The FWHM of the effective spatial resolution is $\sim0\farcs6$. The plus, X, and Y signs indicate the positions where H40$\alpha$ is detected (SECTION \ref{sec:ALMA}). The mapping area is $20"\times20"$ with the center at the northern nucleus (plus sign). The 2D maps are downloaded from the MAD project website \citep{Erroz-Ferrer_2019}}. \label{fig:MUSE}
\end{center}
\end{figure*}

NGC~3256 was observed by MUSE as one of the targets for the MUSE Atlas of Disks (MAD) project \citep{Erroz-Ferrer_2019}.
The processed MUSE 3D data cube of NGC~3256 can be downloaded from the ESO science archive portal\footnote{\url{http://archive.eso.org/scienceportal/home}}, and has a field of view (FoV) of 1 arcmin$^2$, with spatial sampling of 0\farcs2, the full width half maximum (FWHM) of the effective spatial resolution is $\sim0\farcs6$, spectral sampling of 1.25~$\AA$, and an observation date of April 6, 2016.\footnote{ESO Programme ID 097.B-0165.}
Figures~\ref{fig:MUSE} (a) and (b) show the extinction map and extinction-corrected H$\beta$ map processed by \citet{Erroz-Ferrer_2019}. The 2D maps in Figure~\ref{fig:MUSE} are downloaded from the MAD project web-page\footnote{\url{https://www.mad.astro.ethz.ch/data-products}}. We use these 2D maps for the main analysis (i.e., measurements of SFR). 
We use the 3D data cube only for measuring line profiles (see section~\ref{sec:Line}). 
The errors for the emission line flux are about $10\%$ for the low signal-to-noise-ratio (S/N$<3$) regions, and 2$\%$ for the higher S/N regions \citep{Erroz-Ferrer_2019}. A conservative overall photometric error of 5$\%$ is adopted for the analysis using the H$\beta$ map.
In order to compare the MUSE/VLT and ALMA images, the H$\beta$ peak position at the non-dusty northern nucleus is assumed to be same as the H40$\alpha$ peak position.

\subsection{ALMA}\label{sec:ALMA}
The H40$\alpha$, H42$\alpha$, $^{13}$CO~(1--0), and CH$_3$OH~($2_k$--$1_k$) data cubes were obtained as part of the 85--110 GHz range line search ALMA project for NGC~3256 (ID: 2015.1.00993.S).
In addition, the data from two other ALMA projects (ID: 2015.1.00412.S and 2016.1.00965.S) \citep{Harada_2018} were combined during data processing in order to produce higher quality H40$\alpha$ and H42$\alpha$ images (Table~\ref{tab:observation_RL}). 
The calibrated visibility data were obtained by the calibration scripts that were provided by ALMA east Asian Regional Centerand processed using Common Astronomy Software Applications (${\tt CASA}$) \citep{McMullin_2007}.
We manually applied band-edge flagging and flux scaling for the data obtained in one specific Execution Block  ({uid\_\_\_A002\_Xb00ce7\_X47b4}). 
Eight channels were flagged at the band-edge, whereas the original script flags 15 channels which included channels near the H40$\alpha$ emission.
In addition, the absolute flux was corrected by a factor of 1.115 since the continuum flux for this Execution Block was systematically lower than the others. 
The continuum emissions were subtracted using the {\tt uvcontsub} task in ${\tt CASA}$.
The data cubes were produced by using the {\tt tclean} task in ${\tt CASA}$ with the Briggs weighting (robust = 2.0; Natural waiting), the velocity resolution of 50 km s$^{-1}$, and the pixel size of $0\farcs125$.
The clean masks were selected by
the automatic masking loop (sidelobethreshold=2.0, noisethreshold=2.5, lownoisethreshold=1.5, minbeamfrac=0.3, growiterations=75, and negativethreshold=0.0).
The FoV of the ALMA map is $59\farcs$4 at the sky frequency of H40$\alpha$ emission.
For $^{13}$CO~(1--0) imaging, we applied robust = 0.5 since the signal to noise ratio is high enough.
The continuum map was produced using the line-free channels beside the H40$\alpha$ emission line.
Table~\ref{tab:data} is a summary of the achieved angular resolution and sensitivity for each line.

\begin{figure*}
\begin{center}
\includegraphics[width=18cm]{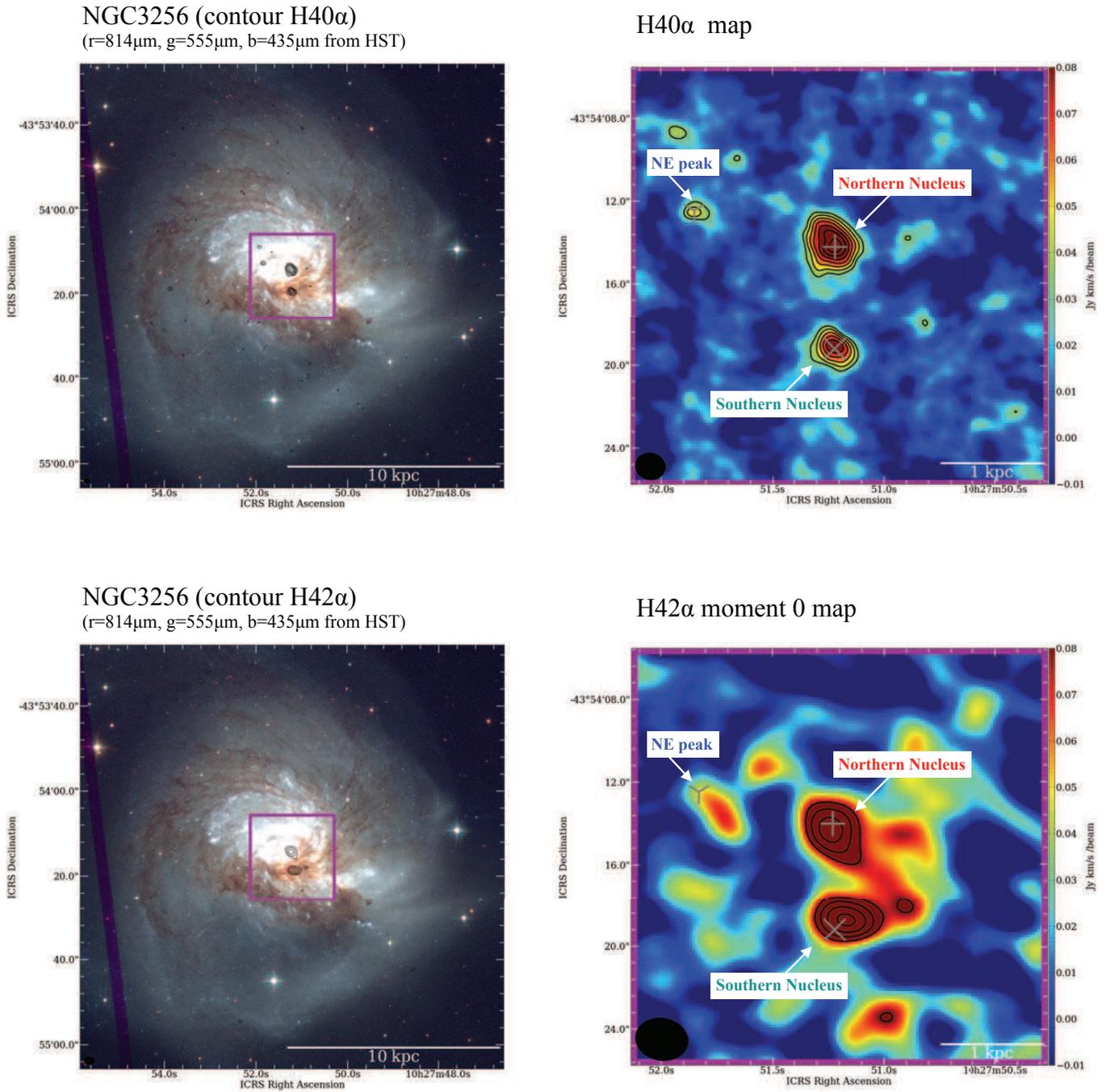}
\caption{
(top left) Optical color image of entire NGC~3256 obtained by HST. 
Black contours show H40$\alpha$ with $11\times(3,4,5,6,7,8,9)$ [mJy beam$^{-1}$ km s$^{-1}$]. A magenta square shows a central $20\arcsec\times20\arcsec$ region.
(top right) Integrated intensity map of H40$\alpha$ image that achieves an angular resolution of $1\farcs48 \times 1\farcs31$. Contours are the same as in the left panel. The plus, X, and Y signs indicate the positions where H40$\alpha$ is detected.
(bottom) Figures as in the top panels for H42$\alpha$ results. Black contours indicate H42$\alpha$ with $22\times(3,4,5,6)$ [mJy beam$^{-1}$ km s$^{-1}$].  Achieved angular resolution is $2\farcs57 \times 2\farcs05$. We detect H42$\alpha$ at the northern and southern nuclei with S/N of $\>5$. 
}
\label{fig:H40a_mom0}
\end{center}
\end{figure*}

\begin{figure*}
\begin{center}
\includegraphics[width=15.0cm]{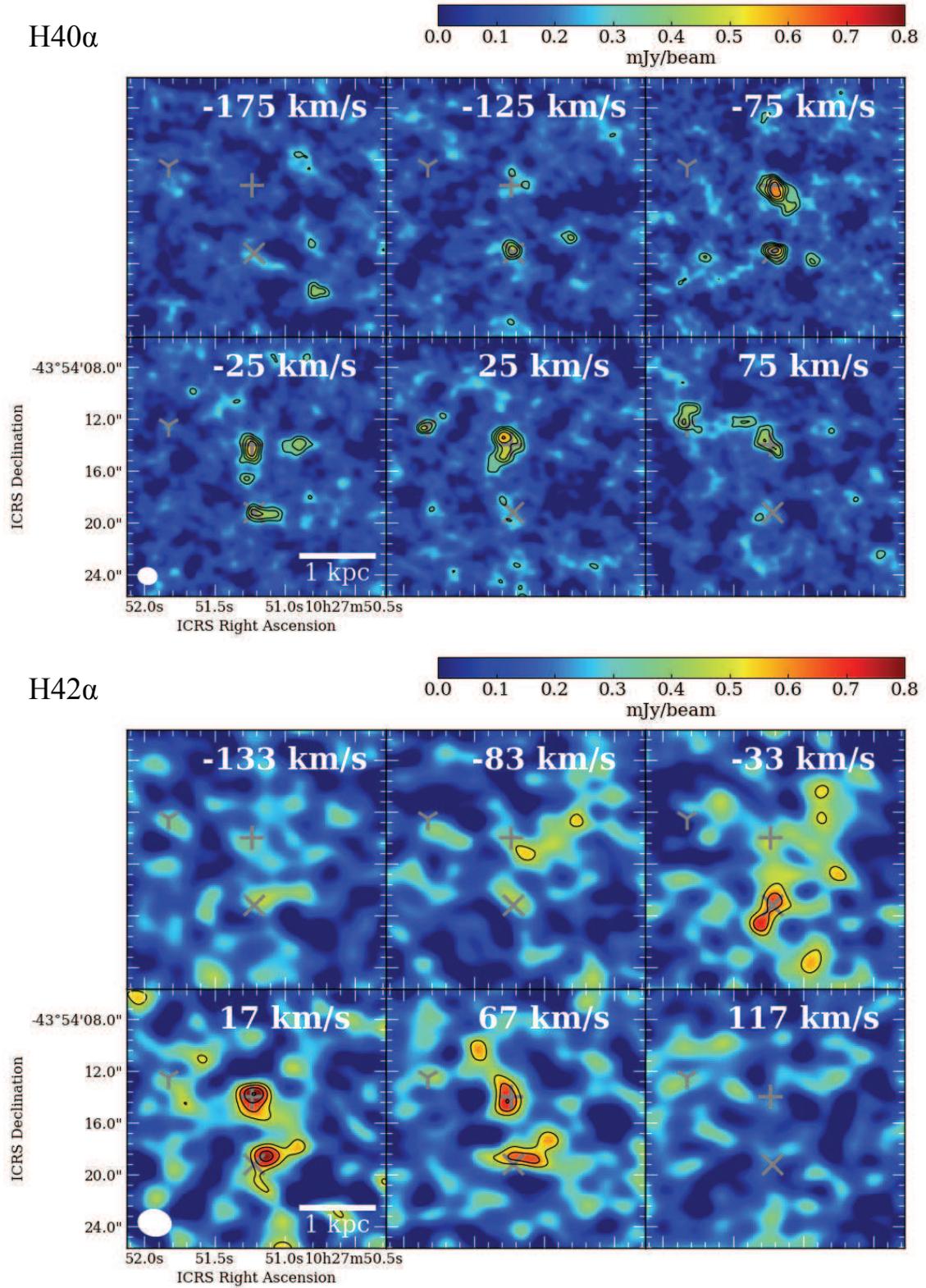}
\caption{
(top) H40$\alpha$ channel map. The contour is $0.08\times(3,4,5,6,7)$ [mJy beam$^{-1}$]. The symbols are the same as those in Figure \ref{fig:MUSE}. The velocity offset from the systematic velocity is labeled.
(bottom) H42$\alpha$ channel map. The contour is $0.18\times(3,4,5)$ [mJy beam$^{-1}$].
}
\label{fig:H40a_chan}
\end{center}
\end{figure*}

\begin{figure*}
\begin{center}
\includegraphics[width=18cm]{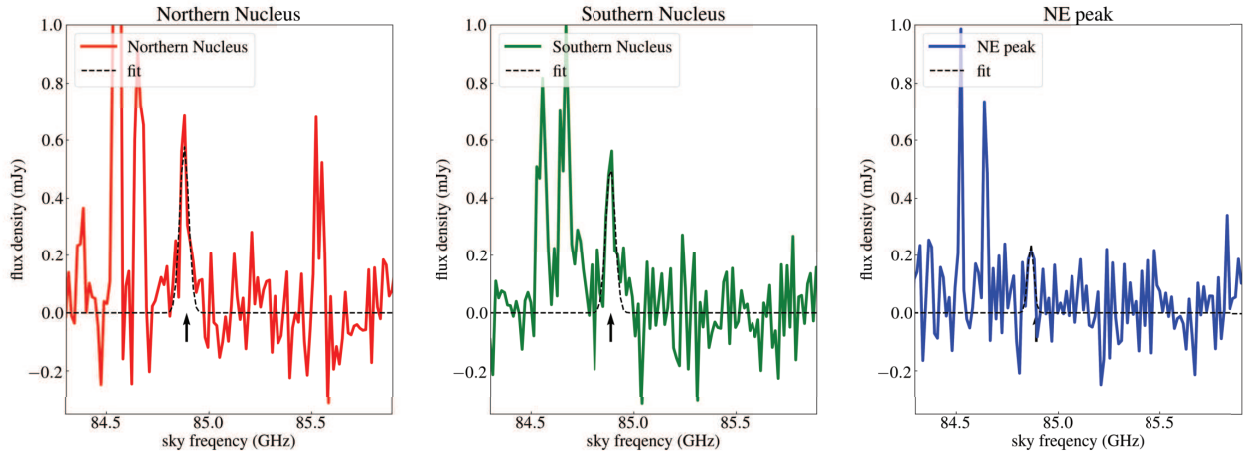}
\caption{
(top) H40$\alpha$ spectrum with photometric beam size of $\sim1\farcs5$ at the northern nucleus (red), southern nucleus (green), and NE peak (blue). The black dashed line is the result of Gaussian fitting. The arrows show the sky frequency at the systematic velocity. (bottom) Same figures for H42$\alpha$. There are three other emission lines in these spectra:
c-C$_3$H$_2$~($2_{1,2}$--$1_{0,1}$) at 84.547 GHz, CH$_3$CCH~(5$_k$--4$_k$) at 84.664~GHz, and SO~(2$_2$--1$_1$) at 85.295 GHz (sky frequency). 
}
\label{fig:H40a_spec}
\end{center}
\end{figure*}

\begin{figure*}
\begin{center}
\includegraphics[width=18.0cm]{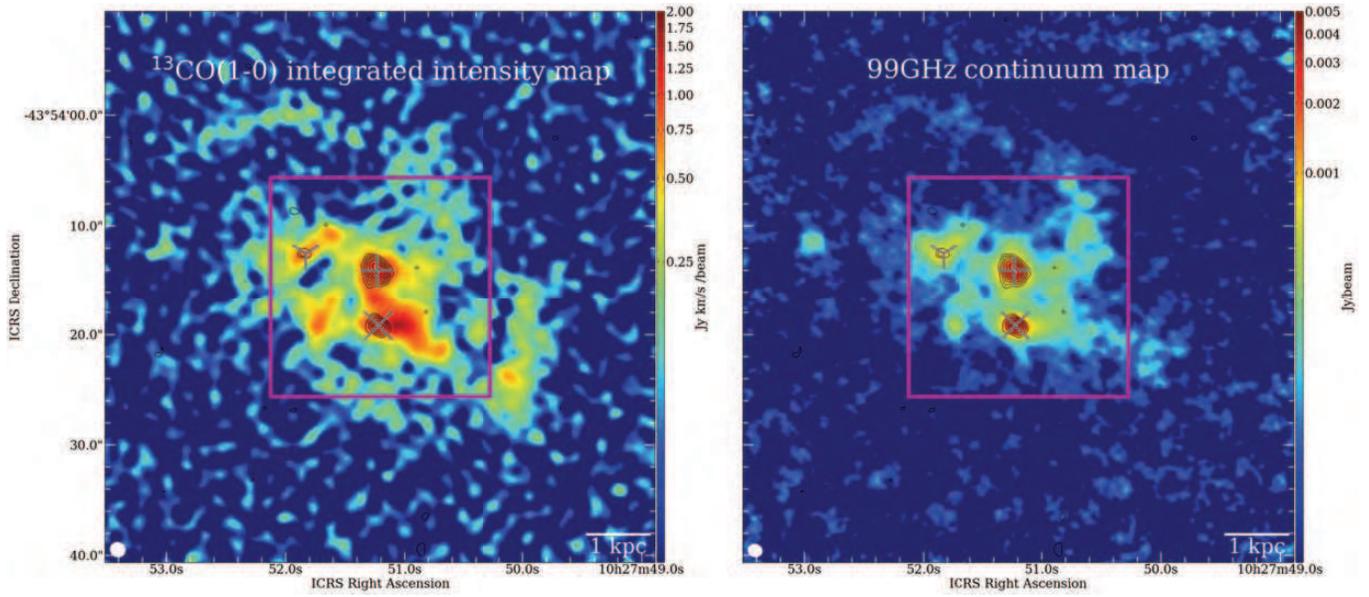}
\caption{
(left) Integrated intensity map of $^{13}$CO~(1--0) emission. (right) Continuum map at rest-frame 99~GHz.
A magenta square shows a central $20\farcs\times20\farcs$ region.
The signs are the same as those in Figure \ref{fig:MUSE}.
}
\label{fig:13co_cont}
\end{center}
\end{figure*}

Figure~\ref{fig:H40a_mom0} shows the Hubble Space Telescope (HST) optical color image\footnote{Based on observations made with the NASA/ESA HST, and obtained from the Hubble Legacy Archive, which is a collaboration between the Space Telescope Science Institute (STScI/NASA), the Space Telescope European Coordinating Facility (ST-ECF/ESA), and the Canadian Astronomy Data Centre (CADC/NRC/CSA). The coordinates are manually corrected based on the positions of stars to compare ALMA images.}
and the integrated intensity map of H40$\alpha$ and H42$\alpha$. 
 Channel maps are shown in Figure~\ref{fig:H40a_chan} and the spectra are shown in Figure~\ref{fig:H40a_spec} for each region.
We use H40$\alpha$ line flux to derive physical parameters, because the image quality (i.e., angular resolution and sensitivity) is better than that of H42$\alpha$.
In order to identify the H~$_{\rm II}$ regions probed by the H40$\alpha$ line, we use the ${\tt imfit}$ task in ${\tt CASA}$ to fit elliptical Gaussian components on the integrated intensity map.
H40$\alpha$ is detected at the northern nucleus, southern nucleus, and northeastern (NE) peak with S/N of $>10$, $>8$, and $>4$, respectively.
Table~\ref{tab:source_ID} is a summary of the coordinates, line flux, and source size (FWHM of major and minor axes) of the detected regions.
Figure~\ref{fig:13co_cont} shows the spatial distribution of $^{13}$CO~(1--0) and 99~GHz continuum emission. 
Disk-wide distributions are seen in both the $^{13}$CO~(1--0) and rest-frame 99~GHz continuum.

\subsection{Line profiles}\label{sec:Line}
\begin{figure}
\begin{center}
\includegraphics[width=9.0cm]{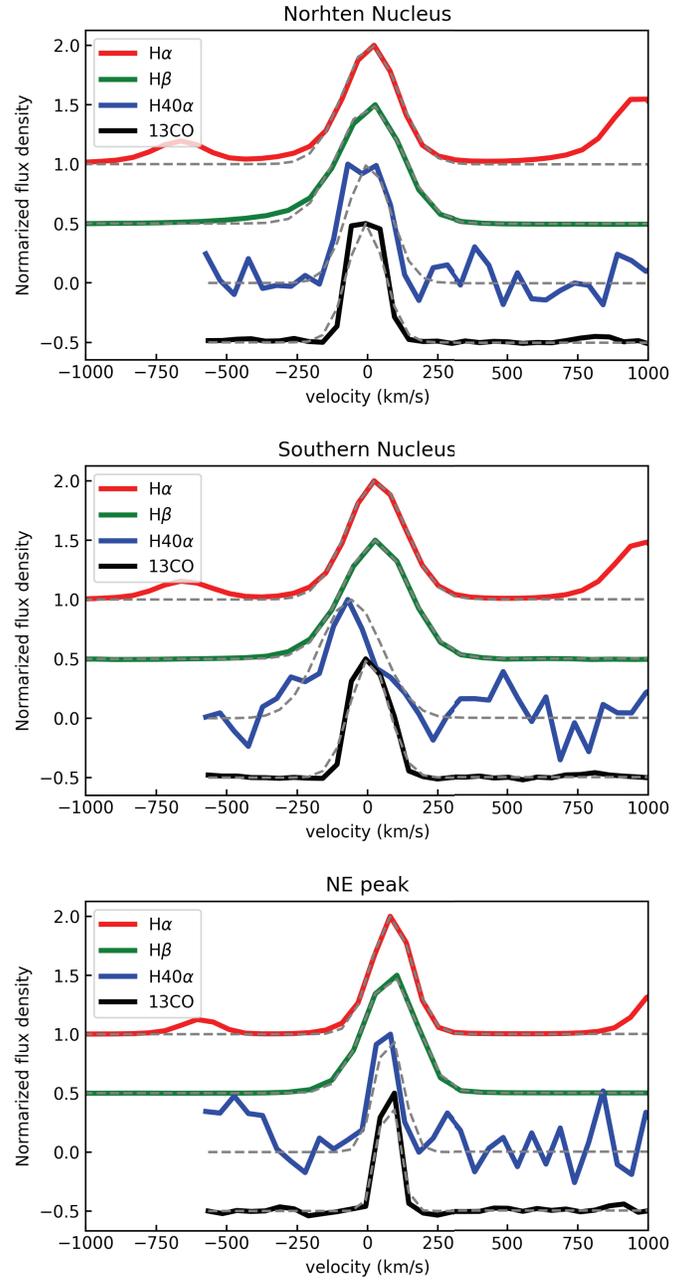}
\caption{Emission line profile for H$\alpha$ (red), H$\beta$ (green), H40$\alpha$ (blue), and $^{13}$CO~(1--0) (black). Gray dashed lines are results for single Gaussian fittings. The spectra are taken at the H40$\alpha$ detected area (Table \ref{tab:source_ID}) at the (top) northern nucleus, (middle) southern nucleus, and (bottom) NE peak.}
\label{fig:profile}
\end{center}
\end{figure}

Figure \ref{fig:profile} shows the line profiles for each line, and the results of Gaussian fittings are shown in Table~\ref{tab:Profile}. We use 3D data cube (without extinction correction) obtained by ESO archive that is not processed by MAD project. The velocity range is consistent among each line. The peak velocity of H40$\alpha$ emission at the southern nucleus is blue-shifted compared with the H$\alpha$ and H$\beta$ lines, while the three lines have similar velocities at  the northern nucleus and NE peak. This may mean that dusty star formation activities that H40$\alpha$ can trace (but optical lines cannot) have different velocity components, yielding variation of the derived SFR between H40$\alpha$ and H$\beta$ lines. The velocity width is larger in the optical lines than in the mm ones. This is likely due to the lower S/N of the H40$\alpha$ detection than optical line detections.

\section{Analysis}\label{Analysis}
\subsection{SFR diagnostic for hydrogen recombination lines}\label{SFR_ALMA}
\begin{figure*}
\begin{center}
\includegraphics[width=9.0cm]{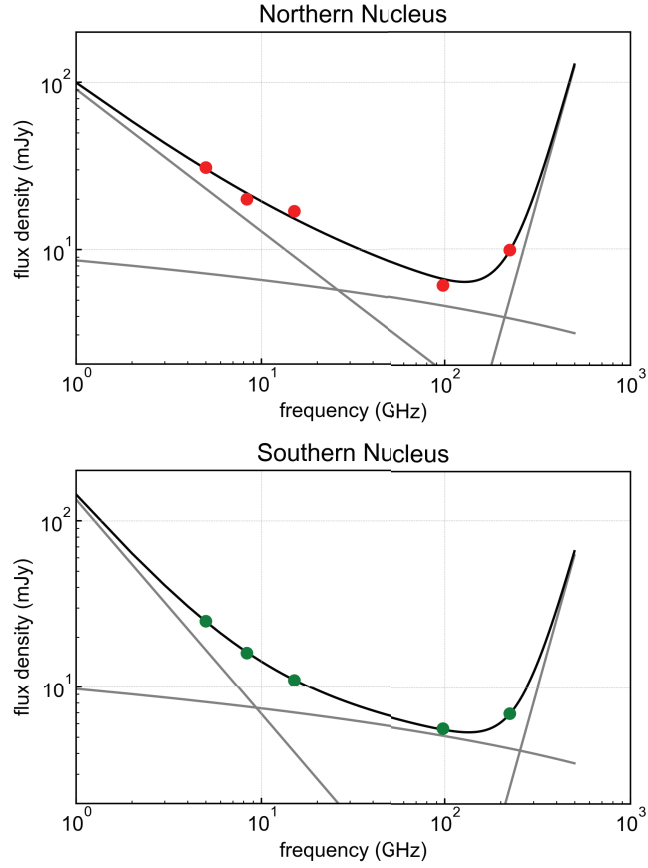}
\caption{
Spectral energy distribution (SED) measured within $2\farcs5$ around the northern nucleus, and $3\farcs5$ around the southern nucleus.
We use the 5.0, 8.3 and 15 GHz continuum flux density measured by Very Large Array (VLA) from \citet{Neff_2003}, and 224~GHz data points from archival ALMA data \citep{Harada_2018}.
The gray lines show the three components.
We use a synchrotron function with a variable power-law index, a free--free function scaled by $g_{\rm FF}$, and a modified Rayleigh--Jeans function (dust) with a fixed index of 4. The calculated synchrotron power-law indexes are -0.85 and -1.29 at the northern and southern nucleus, respectively. The free--free contribution on 99~GHz continuum emission is $\sim76\%$ for the northern nucleus and $\sim90\%$ for the southern nucleus, assuming an electron temperature of $T_{\rm e}=5000~$K.
}
\label{fig:SED}
\end{center}
\end{figure*}

The relation between ionizing photon rate $Q$~[s$^{-1}$] and SFR~[$\rm{M}_{\odot}~\rm{yr}^{-1}$] depends on the initial mass function (IMF), mass range of stellar IMF, and timescale ($\tau$) over which star formation needs to remain constant, and on stellar rotation effects. 
According to \citet{Bendo_2016}, SFR can be calculated using the relation of 
\begin{equation}\label{Q-SFR}
  \frac{SFR}{\rm{M}_{\odot}~\rm{yr}^{-1}} = 5.41\times10^{-54}~\left[\frac{Q}{\rm s^{-1}}\right].
\end{equation}
We note that the coefficient in this equation can vary by a factor of two depending on the adopted assumption (e.g. the coefficient increases without stellar rotation effects). The details are explained in \citet{Bendo_2015, Bendo_2016}.
The emission measure ($EM=n_{\rm e}n_{\rm p}V$, where $n_{\rm e}$, $n_{\rm p}$, and $V$ are ionized electron volume density, proton volume density, and volume of ionized H~$_{\rm II}$ region, respectively) is described using the total recombination coefficient ($\alpha_{\rm B}$).  
\begin{equation}\label{EM-Q}
  \frac{EM}{\rm cm^{-3}} = \left[\frac{Q}{\rm s^{-1}}\right] \left[
\frac{\alpha_{\rm B}}{\rm cm^3~s^{-1}}\right]^{-1}.
\end{equation}
Using the specific emissivity ($\epsilon$) of each recombination line, the recombination line luminosity can be calculated by
\begin{equation}\label{L-EM}
\frac{L_{\rm RL}}{\rm erg~s^{-1}} = \left[\frac{\epsilon}{\rm erg~s^{-1}~cm^{-3}}\right] \left[\frac{EM}{\rm cm^{-3}}\right].
\end{equation}
Here, the emissivity is given per unit $n_{\rm e}n_{\rm p}$.
From equation~ (\ref{Q-SFR})-(\ref{L-EM}),
\begin{equation}\label{SFR-eq}
  \frac{SFR_{\rm RL}}{\rm{M}_{\odot}~\rm{yr}^{-1}} = 
  5.41\times10^{-54}
  \left[\frac{\epsilon}{\rm erg~s^{-1}~cm^{-3}}\right]^{-1}
  \left[\frac{\alpha_{\rm B}}{\rm cm^3~s^{-1}}\right]
  \left[\frac{L_{\rm RL}}{\rm erg~s^{-1}}\right].
\end{equation}
The luminosity can be calculated from the observed total line flux \citep{Solomon_2005}:
\begin{equation}\label{L-FD}
\frac{L_{\rm RL}}{\rm erg~s^{-1}} = 4.0\times10^{30}\left[\frac{\int f_{\nu \rm{RL}}{\rm d}v}{\rm Jy~kms^{-1}}\right] \left[\frac{\nu_{\rm rest}}{\rm GHz} (1+z)^{-1}\right] \left[\frac{D_{\rm L}}{\rm Mpc}\right]^2.
\end{equation}
The calibration constant between SFR$_{\rm RL}$ and recombination line luminosity
is applicable to the case where SFR is constant over $>$6~Myr. 
There is no dependency on long timescales, unlike the calibration constant between SFR and TIR luminosity \citep{Calzetti_2013}.
Finally, the relation between SFR and total line flux \citep{Bendo_2016} is 
\begin{equation}\label{SFR_line}
\begin{split}
\frac{SFR_{\rm RL}}{\rm{M}_{\odot}~\rm{yr}^{-1}}=
2.16\times10^{-23}
\left[\frac{\rm{\alpha_B~cm^6}}{\rm{\epsilon~erg}} \right]
\left[\frac{\rm{\nu}}{\rm{GHz}} \right]
\left[\frac{D_{\rm L}}{\rm{Mpc}} \right]^2
\left[\frac{\int f_{\nu \rm{RL}}{\rm d}v}{\rm{Jy~km~s^{-1}}}\right].
\end{split}
\end{equation}
The $\alpha_{\rm B}$ terms depend on electron temperature ($T_{\rm e}$) and electron density ($n_{\rm e}$), assuming case-B recombination.
The $\alpha_{\rm B}$ values are listed in \citet{Storey_1995}.
We fixed the $n\rm{_e}$ to $10^3$~cm$^{-3}$, as the dependence is negligible in the range of 10$^2$--10$^5$~cm$^{-3}$ \citep{Storey_1995, Bendo_2015}.
We use an interpolated relation between $\alpha_{\rm B}$ and $T_{\rm e}$~(500--30000~K) for hydrogen recombination:
\begin{equation}\label{aB-T}
\frac{\alpha_{\rm B}}{\rm cm^3~s^{-1}} =(3.63\times10^{-10}) \left[\frac{T\rm{_e}}{\rm K}\right]^{-0.79}. 
\end{equation}
The $\epsilon$ values are also listed by \citet{Storey_1995}, and we fixed the $n\rm{_e}$ of $10^3$~cm$^{-3}$ to use the interpolated relation between $\epsilon$ and $T_{\rm e}$ (500--30000 K).
For example, in the case of optical, infrared, and mm recombination lines, 
\begin{eqnarray}\label{ep-T}
 \left[\frac{\epsilon}{\rm erg~s^{-1}~cm^{-3}/{\it n}_{\rm e}{\it n}_{\rm p}}\right] \sim
 \left\{
 \begin{array}{ll}
    (3.00\times10^{-27})~\left[\frac{T\rm{_e}}{\rm K}\right]^{-1.30} & (\rm for~H40\alpha) \\
    (1.22\times10^{-21})~\left[\frac{T\rm{_e}}{\rm K}\right]^{-0.89} & (\rm for~H\alpha) \\
    (1.83\times10^{-22})~\left[\frac{T\rm{_e}}{\rm K}\right]^{-0.80} & (\rm for~H\beta) \\
    (3.36\times10^{-23})~\left[\frac{T\rm{_e}}{\rm K}\right]^{-1.00} & (\rm for~Br\gamma). \\
 \end{array}
 \right.
\end{eqnarray}
In order to estimate the electron temperature, 99~GHz flux density can be used.
The free--free (bremsstrahlung) continuum emission can also be used to probe the ionized gas $EM$.
Therefore, it is possible to calculate SFR from the 99~GHz flux density \citep{Draine_2011, Scoville_2013, Bendo_2016}:
\begin{equation}\label{SFR_cont}
\begin{split}
\frac{SFR\rm{_{cont}}}{\rm{M}_{\odot}~\rm{yr}^{-1}}=
9.49\times10^{10}~
g_{\rm ff}^{-1}
\left[\frac{\rm{\alpha_B}}{\rm{cm^3~s^{-1}}} \right]
\left[\frac{T\rm{_e}}{\rm{K}} \right]^{0.5}
\left[\frac{D_{\rm L}}{\rm{Mpc}} \right]^2
\left[\frac{f_\nu\rm{_{cont}}}{\rm{Jy}}\right],
\end{split}
\end{equation}
\begin{equation}\label{gff}
g_{\rm ff} = 0.5535\ln\left|\left[\frac{T\rm{_e}}{\rm{K}}\right]^{1.5} \left[\frac{\nu}{\rm{GHz}}\right]^{-1} Z^{-1} \right| - 1.682.
\end{equation}
Here, we assume an ionic  charge of $Z=1$.
From equation~ \ref{SFR_line} and \ref{SFR_cont}, the ratio of the line flux density integrated over velocity $v$ to the free--free flux density can be written as
\begin{equation}\label{R_eq}
\begin{split}
R=\frac{\int f_{\nu \rm{RL}}{\rm d}v}{f_\nu\rm{_{cont}}}
\biggl[\frac{\rm{Jy}}{\rm{Jy~km~s^{-1}}} \biggr]
= 4.38\times10^{33}g_{\rm ff}^{-1}~\biggl[ \frac{\rm{\epsilon}}{\rm{erg~s^{-1}~cm^{-3}}} \biggr]
\times\biggl[\frac{\rm{\nu}}{\rm{GHz}} \biggr]^{-1}
\biggl[\frac{T\rm{_e}}{\rm{K}} \biggr]^{0.5}.
\end{split}
\end{equation}
The 99~GHz continuum emission is dominated by free--free emission in most cases \citep{Saito_2016}.
However, there is a possible contribution from non-thermal radio emissions and dust emissions.
In order to check this contribution, we use the 5.0, 8.3, and 15~GHz continuum flux density measured by Very Large Array (VLA) from the literature \citep{Neff_2003} and 200~GHz Band6 data from archival ALMA data \citep{Harada_2018}.
Figure \ref{fig:SED} shows the spectral energy density (SED) of the northern and southern nuclei.
Three components can explain 1--300~GHz SED.
The first component is the power law from non-thermal emission using the slope as a free parameter. The second is the free—free emission, which is scaled by the Gaunt factor (equation \ref{gff}). The third component is dust emission with a slope of 4.0.
Assuming $T\rm{_e}=5000$~K, the SED fittings show that the contribution of free--free emission at 99~GHz continuum flux density (frac-FF) is $\sim76\%$ and  $\sim90\%$ at the northern and southern nuclei, respectively.
We note that the values of frac-FF obtained by SED fitting are not significantly sensitive to the assumption of electron temperature. Therefore, the variation in electron temperature can be investigated by  equation~ \ref{R_eq}.
Subsequently,  $\alpha_B$, $\epsilon$, and SFR can be derived.

\subsection{Molecular gas mass}\label{MH2_ALMA}
Assuming optically thin emission and local thermodynamic equilibrium (LTE) conditions, the molecular gas mass  associated with H40$\alpha$ detected regions can be estimated from $^{13}$CO~(1--0).
It is better to use $^{13}$CO~(1--0) than $^{12}$CO~(1--0) when investigating very dusty regions in LIRGs, because the $^{12}$CO~(1--0) line is most likely optically thick.
It is assumed that the excitation temperature of 10~K \citep{Harada_2018} and the  $^{12}$CO/$^{13}$CO ratios ($R_{12/13}$) of $\sim100$ \citep{Henkel_2014} are constant.
Finally, we use the equation
\begin{equation}\label{MH2}
\frac{M_{\rm H_2}}{M_{\odot}} = 0.41~R_{12/13}~\frac{L'_{\rm ^{13}CO}}{\rm K~km~s^{-1}~pc^2}
\end{equation}
when we derive molecular gas mass from $^{13}$CO luminosity  \citep{Battisti_2014}.
Table~\ref{tab:Gas} shows the information of gas mass in each region.

\subsubsection{Total infrared luminosity}\label{TIR}
The total far infrared luminosity ($L_{\rm TIR}$) of $(4.8\pm0.2)\times10^{11}$~$L_\odot$ (5--1100$~\mu$m) for NGC~3256 is calculated using Spitzer and Hershel observations of 24, 70, 100, 160, and 250~$\mu$m flux density, $S_{24}=12.6\pm0.25$~Jy \citep{Engelbracht_2008}, $S_{70}=120.3\pm6.0$~Jy, $S_{100}=145.4\pm6.8$~Jy, 
$S_{160}=93.48\pm4.68$~Jy,
and $S_{250}=93.48\pm4.68$~Jy \citep{Chu_2017}.
The calibration coefficients derived by \citet{Galametz_2013} are used to calculate $L_{\rm TIR}$. 
The TIR luminosity (8--1000$~\mu$m) calculated using the IRAS flux and coefficients \citep{Sanders_1996, Sanders_2003} is $\sim4.97\times10^{11}$~$L_\odot$. We use the former value of the following sections since the two values are consistent within the error.

\subsection{Results}\label{Res}
The dust-extinction-corrected H$\beta$ map (Figure~\ref{fig:MUSE}b) shows disk-wide starbursts. The total SFR based on the H$\beta$ map is SFR$_{\rm H\beta} ^{\rm total}\sim40\pm2~M_{\odot}~\rm{yr}^{-1}$, assuming $T_{\rm e} = 5000$~K. 
The SFR measured by H$\beta$ is insensitive to $T_{\rm e}$, as the relations of $\alpha_{\rm B}$--$T_{\rm e}$ and $\epsilon$--$T_{\rm e}$ have similar indexes of $\sim 0.8$ (equations \ref{SFR_line}, \ref{aB-T}, and \ref{ep-T}).
However, this value likely underestimates the total SFR due to dust extinction.
Table~\ref{tab:SFR}  shows the SFR at star-forming regions where H40$\alpha$ is detected. 
A conservative overall photometric error of 5$\%$ is adopted\footnote{\url{https://almascience.nrao.edu/documents-and-tools/cycle3/alma-technical-handbook}}.
The SFRs of the three detected regions measured by H40$\alpha$ emissions are 
SFR$_{\rm H40\alpha}^{\rm N}=9.8\pm0.5$, 
SFR$_{\rm H40\alpha}^{\rm S}=6.8\pm0.3$, and 
SFR$_{\rm H40\alpha}^{\rm NE}=0.98\pm0.05~M_{\odot}~\rm{yr}^{-1}$.
In contrast, the SFRs of these regions measured by extinction-corrected H$\beta$ data are
SFR$_{\rm H\beta}^{\rm N}\sim6.8\pm0.3$, 
SFR$_{\rm H\beta}^{\rm S}\sim1.7\pm0.1$, and 
SFR$_{\rm H\beta}^{\rm NE}\sim0.47\pm0.02~M_{\odot}~\rm{yr}^{-1}$.
The systematically lower SFR derived from the H$\beta$ line suggests the presence of intervening dust, especially in the southern nucleus.
Finally, the total SFR (SFR$_{\rm H\beta+H40\alpha}^{\rm total}$) is calculated as SFR$_{\rm H\beta}^{\rm total}-$(SFR$_{(\rm H\beta)}^{\rm N}$ + SFR$_{(\rm H\beta)}^{\rm S}$ + SFR$_{(\rm H\beta)}^{\rm NE})+$(SFR$_{\rm H40\alpha}^{\rm N}$ + SFR$_{\rm H40\alpha}^{\rm S}$ + SFR$_{\rm H40\alpha}^{\rm NE}$) $=40-(6.8+1.7+0.47)+(9.8+6.8+0.98)\sim48\pm2$~$M_{\odot}$~yr$^{-1}$.
We note that the total SFR derived here is estimated assuming all the H$\beta$ and H$\alpha$ emissions originate from H~$_{\rm II}$ regions. As mentioned by \citet{Rich_2011}, shocks could also contribute to the line emissions. 
In order to estimate the regions ionized by pure H~$_{\rm II}$ regions, we use Baldwin, Phillips $\&$ Terlevich (BPT) cuts for each pixel derived by \citet{Erroz-Ferrer_2019}.
The total SFR from pure H~$_{\rm II}$ regions is calculated as $\sim40$~$M_{\odot}$~yr$^{-1}$, which is consistent with SFR$_{\rm H\beta} ^{\rm total}$ derived by this project.
The total SFR from TIR luminosity ($L_{\rm TIR}=(4.8\pm0.2)\times10^{11}$~$L_\odot$) is $51.5\pm2.6$~$M_{\odot}$~yr$^{-1}$, assuming a young starburst (100~Myr), Kroupa IMF, and a mass range of $0.1-100~\rm{M}_{\odot}$ \citep{Calzetti_2013}. The comparison between hydrogen recombination lines and TIR luminosity is investigated in SECTION \ref{timescale} in terms of the starburst age.

\begin{figure}
\begin{center}
\includegraphics[width=9.0cm]{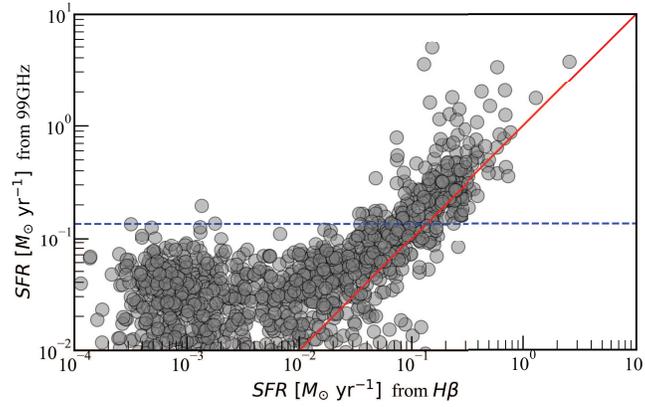}
\caption{Pixel-by-pixel comparison of SFR measured by H$\beta$ and free--free emission. The blue dashed line corresponds to the 3 sigma limits for free--free emissions. The red line shows where the SFR from H$\beta$ and free--free emission is equal. There is a linear relation; however, the SFR traced by free--free emission is systematically higher than that from H$\beta$. This means that 99~GHz emission often has contamination from other properties (e.g., dust and synchrotron).}
\label{fig:Hb_cont}
\end{center}
\end{figure}

Figure~\ref{fig:Hb_cont} shows the relation between SFR derived by H$\beta$ and free--free emission. 
The SFR traced by free--free emission is systematically higher than SFR from H$\beta$, which suggests the contamination from synchrotron and/or dust in the 99~GHz continuum flux density. 
The typical frac-FF can be roughly estimated from the ratio of the SFRs derived by the H$\beta$ and free-free emission.
The mean value of the ratio is $\sim0.7$, indicating the typical frac-FF of $\sim70~\%$. 
This fraction is consistent with the frac-FF of typical starburst galaxies such as NGC~253 \citep{Bendo_2015}.
While uncertainties in the dust extinction correction exist, we adopt the SFR derived using the H$\beta$ line in the following sections because the S/N is higher than the 99GHz continuum map. 
In SECTION~\ref{disk-wide}, we investigate the possible regions where H$\beta$ may underestimate the SFR outside the southern nucleus.

\section{Discussion}\label{Discussion}
The key questions we endeavor to answer are:
(i) `` What is the fraction of star formation missed by optical and infrared observations (e.g., H$\alpha$, H$\beta$, and Br$\gamma$)?";
(ii) `` What is the fraction of the nuclear starburst that contribute to the total SFR?;
(iii) `` Can the variation of gas depletion time be seen within NGC~3256?";
and 
(iv) `` How long is the starburst timescale in NGC~3256?".
 Finally, we investigate the properties of H~$_{\rm II}$ regions (i.e., electron temperature) in H40$\alpha$ detected regions.

\subsection{Northern nucleus}
The northern nucleus contains the largest (area = $0.38\pm0.01$~kpc$^2$) H40$\alpha$ nebula of the three identified (Table~\ref{tab:source_ID}).
The derived SFR is SFR$_{\rm H40\alpha}^{\rm N}=9.8\pm$0.5~$M_\odot$~yr$^{-1}$, which is $\sim20\%$ of the total SFR\footnote{This value is smaller than the SFR derived by IR SED fitting ($\sim15$~$M_\odot$~yr$^{-1}$) \citep{Lira_2008}, which could be due to the different photometric area.}.
The SFR derived from H$\beta$ is  SFR$_{\rm H\beta}^{\rm N}=6.8\pm0.3$~$M_\odot$~yr$^{-1}$, and this is $\sim70\%$ of the SFR derived from H40$\alpha$ (Table~\ref{tab:SFR}).
This difference may be explained by insufficient dust extinction correction which was performed using optical lines alone (i.e., the conversion from  H$\alpha$/H$\beta$ ratio to $A_{\rm V}$).
The star formation rate surface density ($\Sigma^{\rm inner}_{SFR_{\rm H40\alpha}^{\rm {N}}}$) is 32.9$\pm$1.6~$M_\odot$~yr$^{-1}$~kpc$^{-2}$, and the molecular gas mass surface density  ($\Sigma^{\rm inner}_{M_{\rm H_2}}$) around the northern nucleus is $2772\pm139$~$M_\odot$~pc$^{-2}$, which is a typical disk-averaged surface densities for starburst galaxies \citep{Kennicutt_1998}.
This suggests that the characteristics of the  H~$_{\rm II}$ regions near the northern nucleus are consistent with regions in typical starburst galaxies.

\subsection{Southern nucleus}
Despite the significant H40$\alpha$ emission, there is no strong emission in the extinction-corrected H$\beta$ map at the southern nucleus (Figure \ref{fig:MUSE}c).
Consequently, the SFR derived from H40$\alpha$ (SFR$_{\rm H40\alpha}^{\rm S}=6.8\pm$0.3~$M_\odot$~yr$^{-1}$) is larger than that derived from the H$\beta$ map (SFR$_{\rm H\beta}^{\rm S}\sim$$1.75\pm0.09$~$M_\odot$~yr$^{-1}$).
This suggests that the optical emission around the southern nucleus is not originated from extremely dust-obscured nebulae emission; rather, it may be contributed from the different components (e.g., the surface of the dusty star-forming region).
In addition, the offset in the H40$\alpha$ line profile relative to the H$\alpha$ and H$\beta$ lines in the southern nucleus (Table~\ref{tab:Profile} and Fig~\ref{fig:profile}) may be the evidence showing that millimeter and optical lines trace different components.
This demonstrates the benefits of examining both the spectral line parameters as well as the integrated fluxes when investigating dusty starbursts at the nucleus of U/LIRGs.

Emission lines in IR can also be used as an independent proxy of SFR in galaxies.
The southern nucleus can be detectable at wavelength $\gtrsim1\mu$m \citep{Lipari_2000}.
\citet{Piq_2012,Piq_2013} detected Br$\gamma$ emission from the southern nucleus of NGC 3256 (the northern nucleus is not in the FoV) using the Spectrograph for INtegral Field Observations in the Near Infrared (SINFONI) integral field spectroscopy observation with VLT.
We use the Br$\gamma$ data obtained from an online catalog \citep{Piq_2016} and measured the Br$\gamma$ to be $\sim6.3\times10^{-15}$ erg~s$^{-1}$~cm$^{-2}$ at the southern nucleus.
Assuming $A_{\rm Brg}\sim1.2$ estimated from the Br$\gamma$/Br$\delta$ ratio \citep{Piq_2013}, we find that SFR estimated from Br$\gamma$ ($SFR_{({\rm Br\gamma})}$) is $1.4~M_{\odot}$~yr$^{-1}$.
The significantly lower SFR derived from Br$\gamma$ suggests that it may not be an ideal tracer of SFR in dusty regions, such as the southern nucleus of NGC3256. 
The comparison between Br$\gamma$ and H40$\alpha$ flux indicates $A_{\rm Br\gamma}\sim2.4$ ($A_{\rm V}\sim25$ assuming $A_{\rm Br\gamma}=0.096A_{\rm V}$). 

\subsection{Disk-wide starburst}\label{disk-wide}
The sum of the nuclear starbursts in the northern and southern nuclei derived from the H$40\alpha$ data is $16.6\pm0.6~M_\odot$~yr$^{-1}$. 
Using the total SFR of $\sim48\pm2~M_{\odot}~\rm{yr}^{-1}$ (SECTION \ref{Res}), the contributions of the nuclear and disk-wide starbursts are $\sim34~\%$ and $\sim66~\%$, respectively. 
In addition, H40$\alpha$ is detected at NE peak (Figure \ref{fig:H40a_mom0}) on the dust lane of the arm that has offset from the two nuclei.

\begin{figure}
\begin{center}
\includegraphics[width=18.0cm]{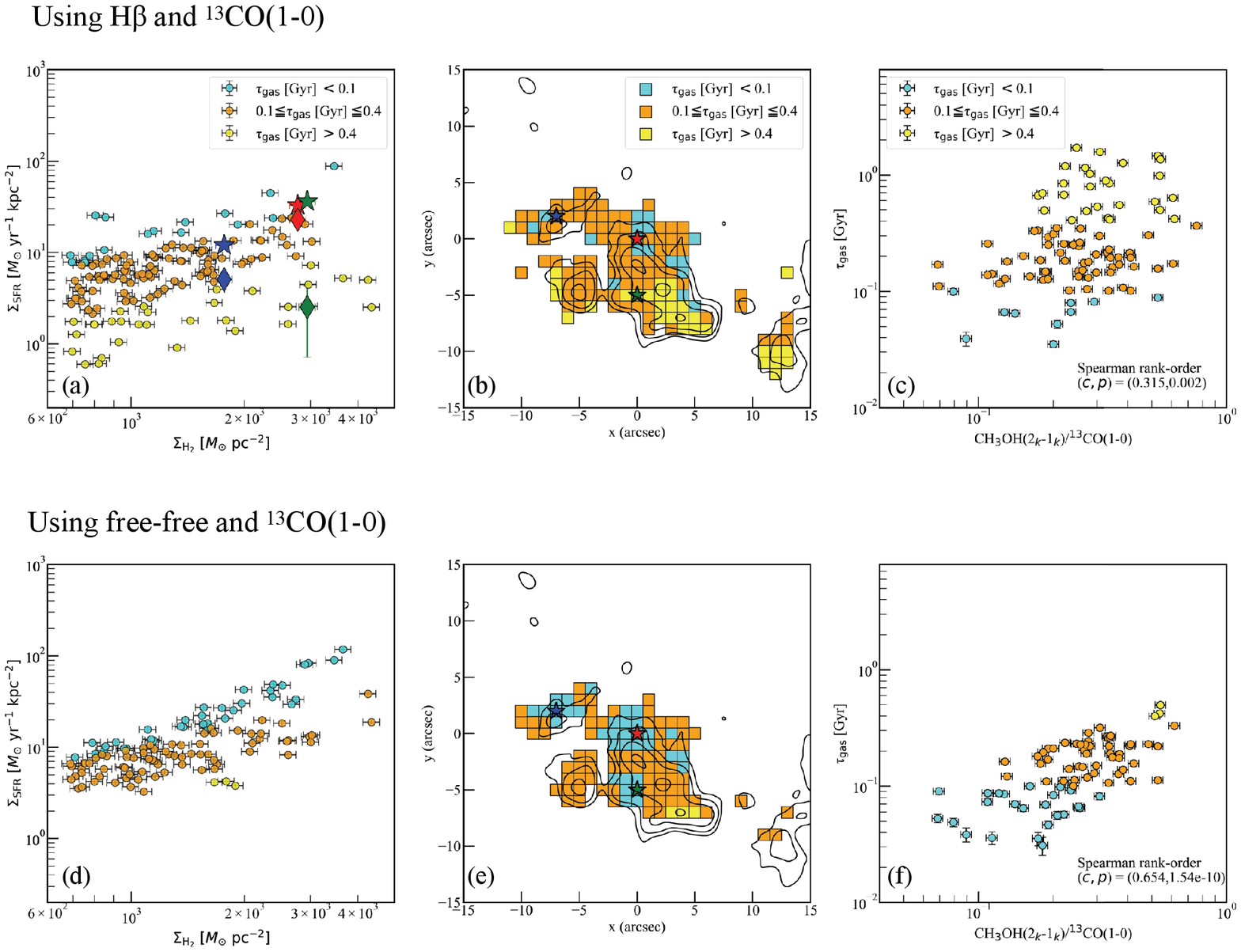}
\caption{(a) Pixel-by-pixel analysis for an empirical star formation relation ($\Sigma_{\rm H_2}-\Sigma_{\rm SFR}$). A pixel corresponds to $1\farcs\times1\farcs$. The red, green, and blue stars indicate the SFR measured by H40$\alpha$ at the northern nucleus, southern nucleus, and NE peak, respectively, in Tables~\ref{tab:SFR} and \ref{tab:Gas}
(The $+$, X, and Y signs in Figures \ref{fig:MUSE} and \ref{fig:H40a_mom0}).
The red, green, and blue diamonds indicate the SFR based on the H$\beta$ map for the northern nucleus, southern nucleus, and NE peak, respectively. The cyan, orange, and yellow circles indicate the pixels with $\tau_{\rm gas}$ [Gyr]$<$0.1, 0.1 $\leqq$ $\tau_{\rm gas}$ [Gyr] $\leqq$ 0.4, $\tau_{\rm gas}$ [Gyr]$>$0.4, respectively. The SFR for each point is calculated using H$\beta$ emission.
(b) Spatial information of $\tau_{\rm gas}$ and CH$_3$OH~(2$_k$--1$_k$). The cyan, orange, and yellow squares indicate the positions for the pixels with  $\tau_{\rm gas}$ [Gyr]$<$0.1, 0.1 $\leqq$ $\tau_{\rm gas}$ [Gyr] $\leqq$ 0.4, $\tau_{\rm gas}$ [Gyr]$>$0.4, respectively. The red, green, and blue stars indicate the respective stars in the panel (a). 
The black contours show the distribution of CH$_3$OH~(2$_k$--1$_k$) at $0.013\times(3,6,12,24)$~Jy~beam$^{-1}$~km~s$^{-1}$. 
(c) Relation between CH$_3$OH~(2$_k$--1$_k$)/$^{13}$~CO~(1--0) and $\tau_{\rm gas}$. The Spearman's rank correlation coefficient and $p$-values are shown. (d)--(f) Same figures as (a)--(c) but with the SFR measured by free--free emission. The SFR from free-free emission is measured by assuming frac-FF $=70~\%$.
}
\label{fig:KS-N3256}
\end{center}
\end{figure}

Figure \ref{fig:KS-N3256}(a) shows that star-forming regions in NGC~3256 have large scatter in the $\Sigma_{\rm H_2}$--$\Sigma_{\rm SFR}$ plane, 
particularly in the regions with $\tau_{\rm gas}$ of $<0.1$~Gyr$^{-1}$ as well as regions with $\tau_{\rm gas}$ of $>0.4$~Gyr outside the nuclear region (the gas depletion time $\tau_{\rm gas}$=$M_{\rm gas}$/SFR). 
This large scatter suggests a non-uniform gas depletion time. 
In addition, from a direct comparison with the results from a broad-band spectral survey of NGC3256 \citep{Harada_2018}, we find that shock gas tracers (e.g., CH$_3$OH, SiO, HNCO) are coincident with the regions where $\tau_{\rm gas}$ is long. 
Figure \ref{fig:KS-N3256}(b) shows the spatial distribution of $\tau_{\rm gas}$ , and the contours show the CH$_3$OH~(2$_k$--1$_k$) emission. 
Figure \ref{fig:KS-N3256}(c) shows the relation between CH$_3$OH~(2$_k$--1$_k$)/$^{13}$CO~(1--0) and $\tau_{\rm gas}$. 
The Spearman's rank correlation coefficient (c-value) is 0.315, possibly suggesting a weak correlation. The probability (p-value) is 0.002, which means that the possibility for rejecting null hypothesis is 2$~\%$.
These suggest that merger-induced large-scale shock can possibly suppress the star formation activity in the disk region, although the statistical significance is not very strong.
Figures~\ref{fig:KS-N3256}(d)--(f)  are similar to (a)--(c) but plotted using the SFR measured by 99~GHz continuum after correcting for the contamination from dust and non-thermal emission (assuming $70~\%$) \citep[see also ][]{Wilson_2019}.
Even after correcting frac-FF,  a few regions have $\tau_{\rm gas}$ $>0.4$~Gyr, suggesting that extinction corrected H$\beta$ map underestimates SFR due to incomplete extinction correction.  Alternatively, frac-FF for 99~GHz continuum is much lower than 70~\%.
It is, however, noteworthy that a possible correlation between CH$_3$OH~(2$_k$--1$_k$)/$^{13}$CO~(1--0) and $\tau_{\rm gas}$ exists (Figure~\ref{fig:KS-N3256}(f)), with  a higher c-value than those shown in Figure~\ref{fig:KS-N3256}(c). 
It is thus necessary to investigate other galaxies for a general conclusion. For example, MUSE/VLT data toward merging starburst galaxies (e.g., VV 114, II ZW 96, IC 214, Arp 256, and NGC 6240) already exist, and future ALMA observations of shocked gas tracers, molecular gas, and $\sim$100 GHz continuum with the same resolution as MUSE/VLT are important to understand whether shocks can indeed suppress star formation activities. 

\subsection{Starburst timescale}\label{timescale}
The calibration constant between SFR and $L_{\rm TIR}$ changes 
depending on how long the currently observed starbursts have remained constant, 
because not only the young stellar population but also old, long-lived, low-mass stars contribute to $L_{\rm TIR}$. 
If the calibration constant is correct, the SFR estimated from $L_{\rm TIR}$ should be the same as SFR$_{\rm RL}$.
Assuming constant star formation and a Kroupa IMF in the stellar mass range of  0.1--100~${M}_{\odot}$,
the ratio of SFR$_{\rm RL}$ to $L_{\rm TIR}$ is calculated as below \citep{Calzetti_2013}:
\begin{equation}\label{SFR_TIR}
 \frac{SFR_{\rm RL}~[M_{\odot}~\rm{yr}^{-1}]}{L_{\rm TIR}~[\rm {erg~s^{-1}}]} =
 \left\{
 \begin{array}{ll}
     1.6\times10^{-44}~&~(\tau=10~{\rm Gyr})\\
     2.8\times10^{-44}~&~(\tau=100~{\rm Myr})\\
     3.7\times10^{-44}~&~(\tau=10~{\rm Myr})\\
 \end{array}
 \right.
\end{equation}
The SFR$_{\rm RL}$ of NGC~3256 is $48\pm2~M_{\odot}~\rm{yr}^{-1}$, 
estimated using the H$\beta$ and H40$\alpha$ maps.
Using this SFR$_{\rm RL}$ and $L_{\rm TIR}=(4.8\pm0.2)\times10^{11}$~$L\odot$ \citep{Sanders_2003}, 
the ratio between SFR$_{\rm RL}$ and $L_{\rm TIR}$ is estimated to be $(2.63\pm0.17)\times10^{-44}$.
This is similar to the theoretical value for $\tau$ = 100~Myr, 
suggesting that the current starburst has continued for $\sim$100~Myr.  
This period is shorter than the age of the merger of NGC~3256 \citep[$\sim$500~Myr;][]{Lipari_2000}.
Thus, it is likely that the current starburst in NGC~3256 was triggered by the galaxy interaction.

\subsection{Electron temperature variations}\label{Te}
The electron temperatures are calculated using equation~(\ref{R_eq}).
The electron temperature around the northern nucleus is 5900$\pm$400K. This value is consistent with H~$_{\rm II}$ regions at the central part  ($<4$~kpc) of the Milky Way \citep{Shaver_1983} and other starburst galaxies (e.g., NGC~253 and NGC~4945) \citep{Bendo_2015,Bendo_2016}. 
In contrast, the  electron temperature around the southern nucleus is $11500^{+800}_{-700}$~K, which is consistent with the H~$_{\rm II}$ regions in the outer part of the Milky Way ($>10$~kpc).
The different electron temperatures between the northern and southern nucleus is originally from the different line ratios of $R=\int f_{\rm \nu line} {\rm d}\nu/f_{\rm \nu cont}$.

The free–free emission flux is comparable between the northern and southern nucleus, while the recombination line flux at the southern nucleus is about half of the northern nucleus.
An empirical relation between electron temperature and metallicity suggests regions with lower metallicity are higher in electron temperature \citep{Shaver_1983}, which is a direct consequence of inefficient cooling in low-metallicity regions \citep{Pagel_1979}. 
Our analysis of NGC~3256 suggest that the metallicity of the extremely dusty ($A_{\rm V}\sim25$) southern nucleus is lower than that of the non-dusty regions where UV and optical emission lines can be detected.
Low-metal environments are seen in other galaxies. For example, \citet{Kewley_2006, Ellison_2013} show that the metallicity in interacting galaxies tends to be lower than in non-interacting systems of equivalent mass, and later \citet{Rupke_2008, Herrera-Camus_2018} find the same trend for U/LIRGs.
The low metallicity at the southern nucleus may suggest the past occurrence of a large-scale inflow of metal-poor gas. 
Other possibilities for the low metallicity include massive outflows \citep{Sakamoto_2014, Michiyama_2018} which can remove gas and metals \citep[e.g.,][]{Chisholm_2018}. 

\subsection{Possible AGN activity}
The origin of recombination line flux may be related to the presence of an AGN, especially at the southern nucleus.  
The presence of an AGN is suggested from the IRAC\footnote{Infrared Array Camera on the Spitzer Space Telescope} color and silicate absorption feature \citep{Ohyama_2015}.
The possible AGN is categorized as a low-luminosity AGN with the 2--10 keV luminosity of $L_{\rm 2-10keV}\sim2\times10^{40}$~erg~s$^{-1}$ \citep{Ohyama_2015,Lehmer_2015}. In order to explain the molecular outflows from the southern nucleus, a previously active AGN is needed \citep{Sakamoto_2014, Michiyama_2018}.
If the AGN ionizes the surrounding gas, the velocity dispersion of hydrogen recombination lines is nominally $>~1000$~km~s$^{-1}$. However, the line profile at the southern nucleus has the same line width as that of the northern nucleus ($\sim300$~km~s$^{-1}$) (Table~\ref{tab:Profile} and Figure~\ref{fig:profile}).
In addition, \citet{Izumi_2016} show that the expected line flux of mm hydrogen recombination lines is too low to be detected even by ALMA.
Therefore, the  H40$\alpha$ emission is likely originated from star formation activity at the southern nucleus. 

The AGN may enhance the total infrared luminosity independent of star formation activities.
In such a case, the expected starburst timescale is shorter than those derived in SECTION~\ref{timescale}. Finally, higher electron temperature in the southern nucleus could be due to previous AGN activities. For example, \citet{Popovic_2003} estimated an electron temperature of $>10,000$~K in broad line regions based on the Boltzmann plot method to Balmer lines, which is higher than typical electron temperature at the typical H~$_{\rm II}$ regions \citep[e.g.,][]{Shaver_1983}.

\section{Summary}\label{Summary-sec}
In order to show evidence of the large contribution of disk-wide starbursts to the total SFR in a merging galaxy NGC 3256, we investigated spatially resolved SFR using optical and mm hydrogen recombination lines. 
At first, we used optical integral field units (MUSE mounted on VLT) to obtain maps of recombination lines  (i.e., H$\alpha$ and H$\beta$). We found many star-forming regions outside the nuclear regions. However, it is difficult to investigate star formation activities in dusty nuclear regions using optical observations. 
ALMA observation of the mm recombination lines H40$\alpha$ and H42$\alpha$ allowed us to the quantify the true star formation activity in these regions. The total SFR obtained by H$\beta$ and H40$\alpha$ line emission is $\sim48\pm2~M_\odot$~yr$^{-1}$.
The main findings are as follows:
\begin{itemize}
\item[(1)] H40$\alpha$ emission is detected at the northern nucleus, southern nucleus,  and NE peak. However, there are no bright H$\beta$ emissions at the southern nucleus. This means that there is a dust-obscured region at the southern nucleus. The SFR from the southern dusty region is $6.8\pm0.3$~$M_\odot$~yr$^{-1}$, which is $\sim14\%$ of the total SFR.
\item[(2)]
The sum of the nuclear starbursts in the northern and southern nuclei is $16.6\pm0.6~M_\odot$~yr$^{-1}$, which means that the contributions of the nuclear and disk-wide starbursts are $\sim34~\%$ and $\sim66~\%$, respectively. 
The disk-wide starbursts are predominant compared to the nuclear starbursts, even considering the very dusty starburst seen in the southern nucleus.
\item[(3)]
We find that $\tau_{\rm gas}$ is not uniform in NGC~3256. 
There are regions with $\tau_{\rm gas}<0.1$~Gyr as well as  regions with $\tau_{\rm gas}>0.4$~Gyr outside the nuclear region. 
One possible explanation is merger-induced large-scale shocks  that suppress star formation activities in the disk region.
\item[(4)]
Recombination lines and total FIR luminosity suggest the current starburst started $\sim$100~Myr ago. This is shorter than the timescale of a merger process ($\sim500$~Myr), and this supports the idea that the current starbursts are triggered by a merger process.
\item[(5)]
The electron temperature is higher in the dusty southern nucleus ($10200^{+700}_{-600}$~K) than in the non-dusty northern nucleus  ($5900^{+400}_{-400}$~K).
One possible explanation is the lower metallicity in the southern nucleus than in the northern nucleus, suggesting metal-poor gas inflows or metal-rich gas outflows at the southern nucleus.
\end{itemize}

\clearpage

\begin{longrotatetable}
\begin{deluxetable*}{ccccc}
\tablecaption{ALMA archive data used in this project \label{tab:observation_RL}}
\tablecolumns{4}
\tablewidth{0pt}
\tablehead{
\colhead{line} &
\colhead{date} &
\colhead{ALMA project ID} & 
\colhead{configuration} &
\colhead{MRS$^{\rm a}$}
}
\startdata
H40$\alpha$ & 2016 March 04 & 2015.1.00993.S  & C36-1/2 & $28\farcs4$ \\
 & 2016 March 07& 2015.1.00993.S  & C36-1/2 & $24\farcs7$  \\
 & 2016 May 01 & 2015.1.00412.S  & C36-2/3  & $24\farcs7$ \\
 & 2016 May 28 & 2015.1.00412.S  & C40-3 & $26\farcs6$\\
 & 2016 May 29 & 2015.1.00412.S  & C40-3 & $26\farcs6$\\
 & 2016 October 29 & 2016.1.00965.S  & C40-6 & $3\farcs1$\\
H42$\alpha$ & 2016 March 04& 2015.1.00993.S  & C36-1/2 & $28\farcs2$  \\
 & 2016 March 07& 2015.1.00993.S  & C36-1/2 & $24\farcs7$  \\
$^{13}$CO~(1--0) & 2016 March 09& 2015.1.00993.S  & C36-1/2 & $24\farcs7$ \\
 & 2016 March 11 & 2015.1.00993.S  & C36-1/2 & $24\farcs7$ \\
 CH$_3$OH~($2_k$--$1_k$)  & 2016 March 09& 2015.1.00993.S  & C36-1/2 & $24\farcs7$ \\
 & 2016 March 11 & 2015.1.00993.S  & C36-1/2  & $24\farcs7$\\
 &  2016 May 01 & 2015.1.00412.S  & C36-2/3 & $26\farcs6$ \\
\enddata
\tablenotetext{a}{Maximum recovery scale (MRS) for each configuration at 100~GHz according to ALMA configuration schedule (\url{https://almascience.nrao.edu/observing/observing-configuration-schedule})}
\end{deluxetable*}
\end{longrotatetable}

\begin{longrotatetable}
\begin{deluxetable*}{cccccccccc}
\tablecaption{Achieved angular resolution and sensitivity for ALMA data \label{tab:data}}
\tablecolumns{5}
\tablewidth{0pt}
\tablehead{
\colhead{line} &
\colhead{frequency (rest / sky)} &
\colhead{robust} &
\colhead{beamsize (P.A.)} & 
\colhead{rms$^{\rm a}$} & 
\colhead{rms} &
\colhead{rms} \\
\colhead{} &
\colhead{} &
\colhead{} &
\colhead{} &
\colhead{cube} &
\colhead{integrated intensity} &
\colhead{continuum} \\
\colhead{} & 
\colhead{GHz} &
\colhead{} & 
\colhead{\arcsec} &
\colhead{mJy~beam$^{-1}$} & 
\colhead{mJy~beam$^{-1}$~km~s$^{-1}$} &
\colhead{mJy~beam$^{-1}$}
}
\startdata
H40$\alpha$ &(99.02 / 98.10) & 2.0 & 1.48~$\times$~1.31~($79^\circ$) & 0.08 & 11 & --\\
H42$\alpha$ &(85.69 / 84.89) &2.0 & 2.57~$\times$~2.05~($80^\circ$) & 0.18 & 27 & --\\
$^{13}$CO~(1--0) &(110.20 / 109.18) &  0.5 & 1.43~$\times$~1.33~($56^\circ$) & 0.24 & 50 & --\\
CH$_3$OH~($2_k$--$1_k$) &($\sim$96.74 / $\sim$95.84) &  2.0 & 1.88~$\times$~1.77~($66^\circ$) & 0.11 & 15 & --\\
continuum &(98.61 / 97.70) & 2.0 & 1.32~$\times$~1.20~($77^\circ$)  & -- & -- & 0.028\\
\enddata
\tablenotetext{a}{The velocity resolution is 50~km~s$^{-1}$.}
\end{deluxetable*}
\end{longrotatetable}

\begin{longrotatetable}
\begin{deluxetable*}{cccccccc}
\tablecaption{H40$\alpha$ source identification \label{tab:source_ID}}
\tablecolumns{4}
\tablewidth{1000pt}
\tablehead{
\colhead{position} &
\colhead{R.A. (ICRS)} &
\colhead{Dec. (ICRS)} & 
\colhead{Peak} & 
\colhead{major$^{\rm a}$} &
\colhead{minor$^{\rm a}$} &
\colhead{PA} &
\colhead{area}  \\
\colhead{} &
\colhead{} &
\colhead{} & 
\colhead{Jy~km~s$^{-1}$~beam$^{-1}$} & 
\colhead{\arcsec} &
\colhead{\arcsec} &
\colhead{$^\circ$} &
\colhead{kpc$^2$}
}
\startdata
Northern nucleus & $10^{\rm{h}} 27^{\rm{m}} 51\fs22$ & $-43\degr 54\arcmin 17\farcs$85  & 93 & 2.20$\pm$0.17 & 1.90$\pm$0.13 & 169 & 0.095$\pm$0.010 \\
Southern nucleus & $10^{\rm{h}} 27^{\rm{m}} 51\fs22$ & $-43\degr 54\arcmin 19\farcs$20  & 85 & 1.79$\pm$0.15 & 1.55$\pm$0.11 & 50 & 0.063$\pm$0.007 \\
NE peak & $10^{\rm{h}} 27^{\rm{m}} 51\fs85$ & $-43\degr 54\arcmin 12\farcs46$  & 44 & 1.63$\pm$0.26 & 1.27$\pm$0.16 & 41 & 0.047$\pm$0.010 \\
\enddata
\tablenotetext{a}{The source sizes are measured as the half-light radius for 2D Gaussian fitting.}
\end{deluxetable*}
\end{longrotatetable}

\begin{longrotatetable}
\begin{deluxetable*}{lccc}
\tablecaption{Line profiles \label{tab:Profile}}
\tablecolumns{5}
\tablewidth{1100pt}
\tablehead{
\colhead{position} &
\colhead{line} &
\colhead{velocity offset} &
\colhead{$FWHM$}\\
\colhead{} &
\colhead{} &
\colhead{km~s$^{-1}$} &
\colhead{km~s$^{-1}$}
}
\startdata
Northern nucleus & H$\alpha$ & -10 & 280 \\ 
Northern nucleus & H$\beta$ & -10 & 310 \\ 
Northern nucleus & H40$\alpha$ & 0 & 180 \\ 
Northern nucleus & $^{13}$CO~(1--0) & -20 & 140 \\ 
Southern nucleus & H$\alpha$ & 30 & 260 \\ 
Southern nucleus & H$\beta$ & 30 & 280 \\ 
Southern nucleus & H40$\alpha$ & -60 & 250 \\ 
Southern nucleus & $^{13}$CO~(1--0) & 20 & 170 \\ 
NE peak & H$\alpha$ & 80 & 170 \\ 
NE peak & H$\beta$ & 80 & 220 \\ 
NE peak & H40$\alpha$ & 70 & 120 \\ 
NE peak & $^{13}$CO~(1--0) & 70 & 90 \\ 
\enddata
\tablenotetext{}{The MUSE spectral resolution of 1.25$\AA$ corresponds to the velocity resolutions of $\sim57$~km~s$^{-1}$ for H$\alpha$ and $\sim77$~km~s$^{-1}$ for H$\beta$. In the case of ALMA observations, the velocity resolution is 50~km~s$^{-1}$.}
\end{deluxetable*}
\end{longrotatetable}

\begin{longrotatetable}
\begin{deluxetable*}{lccc}
\tablecaption{Molecular gas mass derived by $^{13}$CO~(1--0) \label{tab:Gas}}
\tablecolumns{5}
\tablewidth{2500pt}
\tablehead{
\colhead{position} &
\colhead{$ \int f_\nu\rm{_{13CO}~d\nu}$} &
\colhead{$M_{\rm H_2}$} & 
\colhead{$\Sigma_{M_{\rm H2}}$}\\
\colhead{} &
\colhead{Jy~km~s$^{-1}$} &
\colhead{$10^8~M_\odot$} & 
\colhead{M$_\odot$~pc$^{-2}$}
}
\startdata
Northern nucleus & 6.06$\pm$0.3 & 14.2$\pm$0.7 & 1756$\pm$88 \\ 
Southern nucleus & 4.79$\pm$0.24 & 11.2$\pm$0.6 & 2091$\pm$105 \\ 
NE peak & 0.66$\pm$0.03 & 1.5$\pm$0.1 & 1666$\pm$83 \\
\enddata
\tablenotetext{}{The flux is measured by the same aperture as in Table~\ref{tab:SFR}. If we use the source size in Table~\ref{tab:source_ID}, the surface density $\Sigma_{M_{\rm H2}}^{\rm inner}$ = $2772\pm139$, $2936\pm147$, and $1769\pm88$~M$_\odot$~pc$^{-2}$ for the northern nucleus, southern nucleus, and NE peak, respectively.}
\end{deluxetable*}
\end{longrotatetable}

\begin{longrotatetable}
\begin{deluxetable*}{lcccccccccc}
\tablecaption{SFR derived by recombination lines \label{tab:SFR}}
\tablewidth{1000pt}
\tablehead{
\colhead{position} &
\colhead{$ \int f_\nu\rm{_{H40\alpha}~d\nu}^{\rm a}$} &
\colhead{$f_\nu\rm{_{cont}}^{\rm b}$} &
\colhead{$R$} &
\colhead{$T_{\rm e}$} &
\colhead{$SFR_{\rm H40\alpha}$} &
\colhead{$\Sigma_{SFR_{\rm H40\alpha}}$ $^{\rm c}$} &
\colhead{$\int f_\nu\rm{_{H\beta}~d\nu}$} &
\colhead{$SFR_{{\rm H\beta}}$} \\
\colhead{} &
\colhead{mJy~km~s$^{-1}$} &
\colhead{mJy} &
\colhead{km~s$^{-1}$} &
\colhead{K} &
\colhead{M$_\odot$~yr$^{-1}$} &
\colhead{M$_\odot$~yr$^{-1}$~kpc$^{-2}$} &
\colhead{10$^{-13}$~erg~s$^{-1}$~cm$^{-2}$} &
\colhead{M$_\odot$~yr$^{-1}$}} 
\startdata 
Northern Nucleus & 273$\pm$14 & 6.36$\pm$0.32 & 43$\pm$3 & $5900^{+400}_{-400}$ & 9.8$\pm$0.5 & 12.1$\pm$0.6 & 28.6$\pm$1.4 & 6.82$\pm$0.34 \\ 
Southern Nucleus & 143$\pm$7 & 5.96$\pm$0.30 & 24$\pm$2 & $10200^{+700}_{-600}$ & 6.8$\pm$0.3 & 12.7$\pm$0.6 & 7.3$\pm$0.4 & 1.75$\pm$0.09 \\ 
NE peak & 26$\pm$1 & 0.66$\pm$0.03 & 40$\pm$3 & $6300^{+400}_{-400}$ & 0.98$\pm$0.05 & 10.7$\pm$0.5 & 2.0$\pm$0.1 & 0.47$\pm$0.02 \\ 
\enddata
\tablenotetext{a}{The flux in this table is measured by an aperture 2.5 times larger than the source size shown in Table~\ref{tab:source_ID} to cover all the flux from the northern and southern nuclei. However, the H40$\alpha$ emission at the NE peak is not spatially resolved, and an aperture 1.2 times larger than the source size is applied.}
\tablenotetext{b}{The 99~GHz continuum flux is corrected using frac-FF of $76\%$,  $90\%$, and  $80\%$ at the northern nucleus, southern nucleus, and NE peak, respectively.}
\tablenotetext{c}{The surface density is measured by the aperture shown in table note a. If we measure the surface density inside the source size shown in Table~\ref{tab:source_ID},  $\Sigma_{SFR_{\rm H40\alpha}}^{\rm inner} = 32.9\pm1.6$, $36.2\pm1.8$, and $12.0\pm0.3$~M$_\odot$~yr$^{-1}$~kpc$^{-2}$ for the northern nucleus, southern nucleus, and NE peak, respectively.}
\end{deluxetable*}
\end{longrotatetable}

\clearpage
\acknowledgments
This work was supported by the National Science Foundation of China (11721303, 11991052) and the National Key R\&D Program of China (2016YFA0400702).
This work was supported in part by the Center for the Promotion of Integrated Sciences (CPIS) of SOKENDAI.
T.M. was financially supported by a Research Fellowship from the Japan Society for the Promotion of Science (JSPS) for Young Scientists. T.M. is supported by JSPS KAKENHI grant No. 18J11194. 
D.I. is supported by JSPS KAKENHI grant No.15H02074.
K.N. is supported by JSPS KAKENHI grant No.15K05035 and No. 19K03937.
This paper makes use of the following ALMA data: ADS/JAO.ALMA $\#$2015.1.00993.S, $\#$2015.1.00714.S, and $\#$2015.1.00412.S.
ALMA is a partnership of ESO (representing its member states), NSF (USA) and NINS (Japan), together with NRC (Canada), MOST and ASIAA (Taiwan), and KASI (Republic of Korea), in cooperation with the Republic of Chile. The Joint ALMA Observatory is operated by ESO, AUI/NRAO and NAOJ.
This research has made use of the NASA/IPAC Extragalactic Database (NED), which is operated by the Jet Propulsion Laboratory, California Institute of Technology, under contract with the National Aeronautics and Space Administration. This research has made use of NASA's Astrophysics Data System. 
The authors also thank Masato Onodera for useful comments about MUSE/VLT data products.
We are grateful to the anonymous referee for useful comments which helped the authors to improve the paper.

\setcounter{section}{0}
\renewcommand{\thesection}{\Alph{section}}
\setcounter{figure}{0}
\renewcommand{\thefigure}{\Alph{section}.\arabic{figure}}

\end{document}